\documentclass[12pt]{article}

\advance\voffset by -0.25cm
\usepackage[utf8x]{inputenc}
\usepackage[round]{natbib}

\usepackage{amsmath,amssymb}
\usepackage{bbm}
\usepackage{graphicx}
\usepackage{longtable,dcolumn,booktabs}
\usepackage{textcomp}

\def\calN{\mathcal{N}}
\def\calP{\mathcal{P}}
\def\calS{\mathcal{S}}
\def\calT{\mathcal{T}}

\newcommand{\E}{\ensuremath{\mathrm{E}}}

\newcommand{\Cov}{\ensuremath{\mathrm{Cov}}}

\def\indicator#1{\mathbbm{1}_{\{#1\}}}

\begin{document}

\title{Spatially adaptive post-processing of ensemble forecasts for temperature}

\author{Michael Scheuerer, Luca B\"uermann}

\maketitle


\begin{abstract}
 We propose an extension of the non-homogeneous Gaussian regression (NGR) model by \citet{Gneiting&2005} that yields locally calibrated probabilistic forecasts of temperature, based on the output of an ensemble prediction system (EPS). Our method represents the mean of the predictive distributions as a sum of short-term averages of local temperatures and EPS-driven terms. For the spatial interpolation of temperature averages and local forecast uncertainty parameters we use a Gaussian random field model with an intrinsically stationary component that captures large scale fluctuations and a location-dependent nugget effect that accounts for small scale variability. Based on the dynamical forecasts by the COSMO-DE-EPS and observational data over Germany we evaluate the performance of our method and and compare it with other post-processing approaches such as geostatistical model averaging. Our method yields locally calibrated and sharp probabilistic forecasts and compares favorably with other approaches. It is 
reasonably simple, computationally efficient, and therefore suitable for operational usage in the post-processing of temperature ensemble forecasts. 
\end{abstract}


\section{Introduction}
\label{intro}
The introduction of ensemble prediction systems marks a radical change in the practice of numerical weather prediction. Traditionally, deterministic forecasts of future states of the atmosphere have been obtained by discretizing the system of partial differential equations that represent the physics of the atmosphere and running them forward in time, starting from initial conditions that describe the current state of the atmosphere on the discretization grid. Due to the non-linearity of the system, uncertainties in the initial conditions (incomplete network of observations, etc.) and in the model formulation lead to fast growing forecast errors. In order to represent the corresponding forecast uncertainty, ensemble prediction systems generate several different forecasts of the same weather variable by perturbing initial conditions and model parameters \citep{Toth&2001}. Combinations of ensemble member forecasts are often more accurate than any of these forecasts individually \citep{Palmer2002}, and by 
interpreting them as a sample of a predictive distribution weather forecasts become probabilistic.

The objective of probabilistic forecasting should be the maximization of sharpness subject to calibration \citep{GneitingBalabdaouiRaftery2007}, which implies in particular that the variance of the predictive distribution should be small but must reflect the true forecast uncertainty. Measures of ensemble spread can be skillful indicators of prediction accuracy \citep{Scherrer&2004}, but in general forecast ensembles cannot accommodate all sources of uncertainty and are often underdispersive \citep{HamillColucci1997}. Statistical post-processing techniques which recalibrate the forecast ensemble have therefore become an important part of any ensemble prediction system \citep{GneitingRaftery2005}. A common idea behind all such methods is the use of information from past forecast-observation pairs for adjusting future forecasts. In this article, we consider methods that transform the ensemble forecasts into a full predictive cumulative distribution function (CDF) by fitting a parametric model to the training 
data. For temperature, the weather variable on which we will focus in this article, several such methods have been proposed in the literature. The ensemble model output statistics (EMOS) method \citep{Gneiting&2005} uses a Gaussian predictive distribution with mean represented by a (positive) linear combination of the ensemble member forecasts plus intercept, and variance equal to a constant offset plus the scaled ensemble variance. It is also referred to as non-homogeneous Gaussian regression (NGR). Bayesian model averaging \citetext{BMA, \citealp{Raftery&2005}} associates each ensemble member with a Gaussian distribution and forms the predictive distribution as a mixture of these member distributions with weights that reflect each member's skill. Various extensions of these methods exist, including a fully Bayesian approach by \citet{DiNarzoCocchi2010}.

Over sufficiently homogeneous regions, it is reasonable to use the same parameters for all forecast locations. Some weather variables, however, depend on factors that are quite variable in space so that the assumption of spatial homogeneity is questionable. Temperature strongly depends on altitude, but even if altitude-related effects are accounted for different regions can have very different characteristics. A recent paper presenting an evaluation of the multimodel EPS 'GLAMEPS' \citep{Iversen&2011} concludes that statistical post-processing ``improves reliability, but needs further elaboration to account for geographical variations''. The obvious solution, fitting a different set of parameters to every observation site, poses the challenge of extrapolating these parameters to locations where forecasts are desired, but no station data for calibration is available. For model output statistics techniques \citep{GlahnLowry1972}, which have traditionally been used for bias correction of numerical weather 
predictions (NWP), the related problem of bias removal across the entire model grid based on observational data has been addressed by several authors. In order to interpolate the biases, initially only known at observation locations, \citet{HackerRife2007} use an interpolation scheme akin to the minimum-variance state estimate in the data assimilation problem, while \citet{Mass&2008} and \citet{Glahn&2009} propose bias correction strategies which explicitly account for elevation, land use type, etc. In the context of ensemble post-processing, \citet{Kleiber&a2011} present an approach called geostatistical model averaging (GMA) which is based on BMA and uses techniques from geostatistics to interpolate post-processing parameters estimated at observation sites to the model grid. The method presented in this article is similar, but builds upon the conceptually simpler EMOS approach. It uses a parametrization of the predictive CDF that is slightly different from \citet{Gneiting&2005}, representing its mean as 
the sum of the short-term average of local temperatures, and EPS-driven terms that account for deviations from this average. The temperature average component naturally entails local adaptivity of the forecast mean, while a location-dependent predictor variable accounts for spatial variations of the associated forecast uncertainty.

\begin{figure}[t]
 \centering
 \includegraphics[width=.48\textwidth]{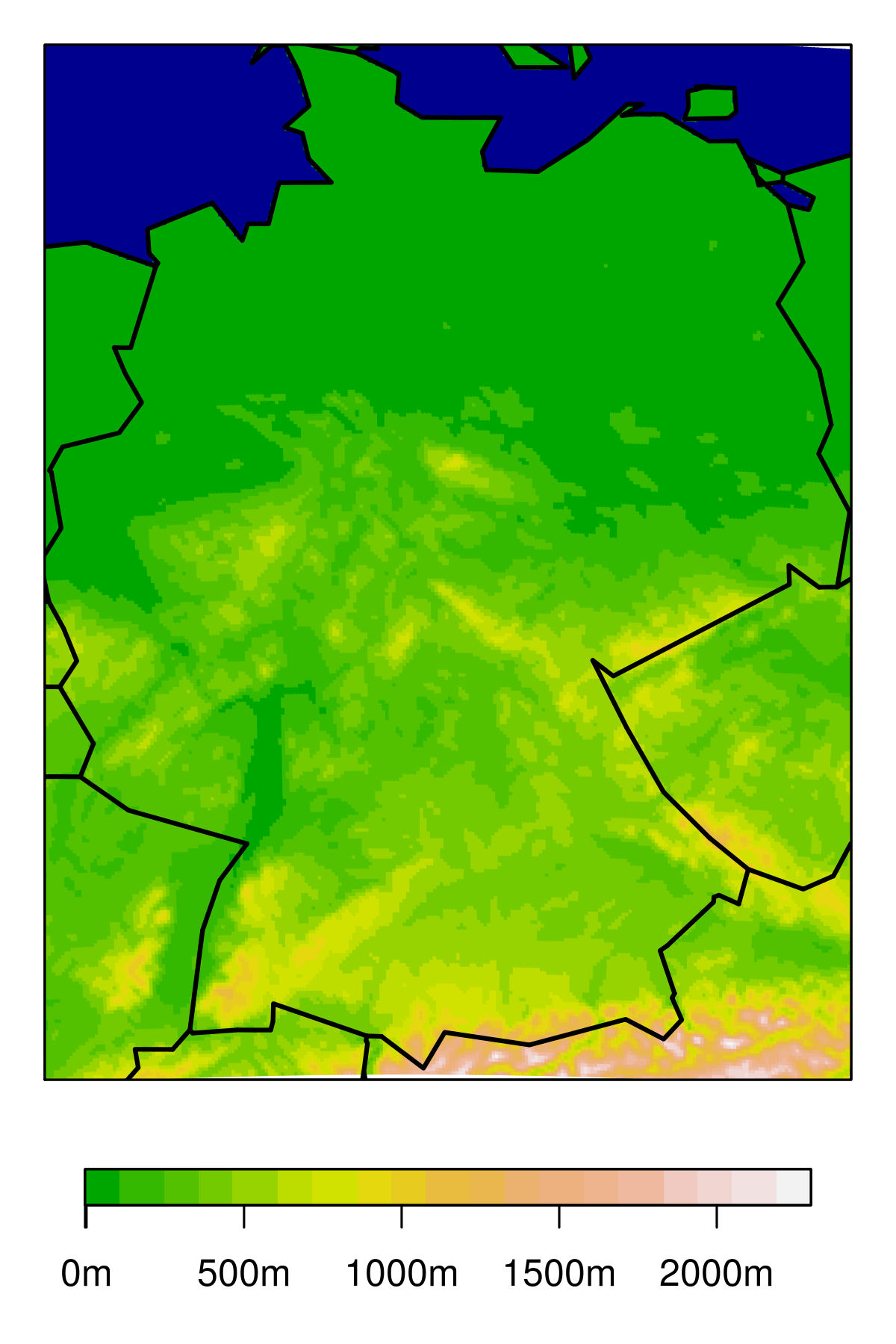}
 \hspace{0.2cm}
 \includegraphics[width=.46\textwidth, trim=1cm 0cm 3cm 2cm, clip]{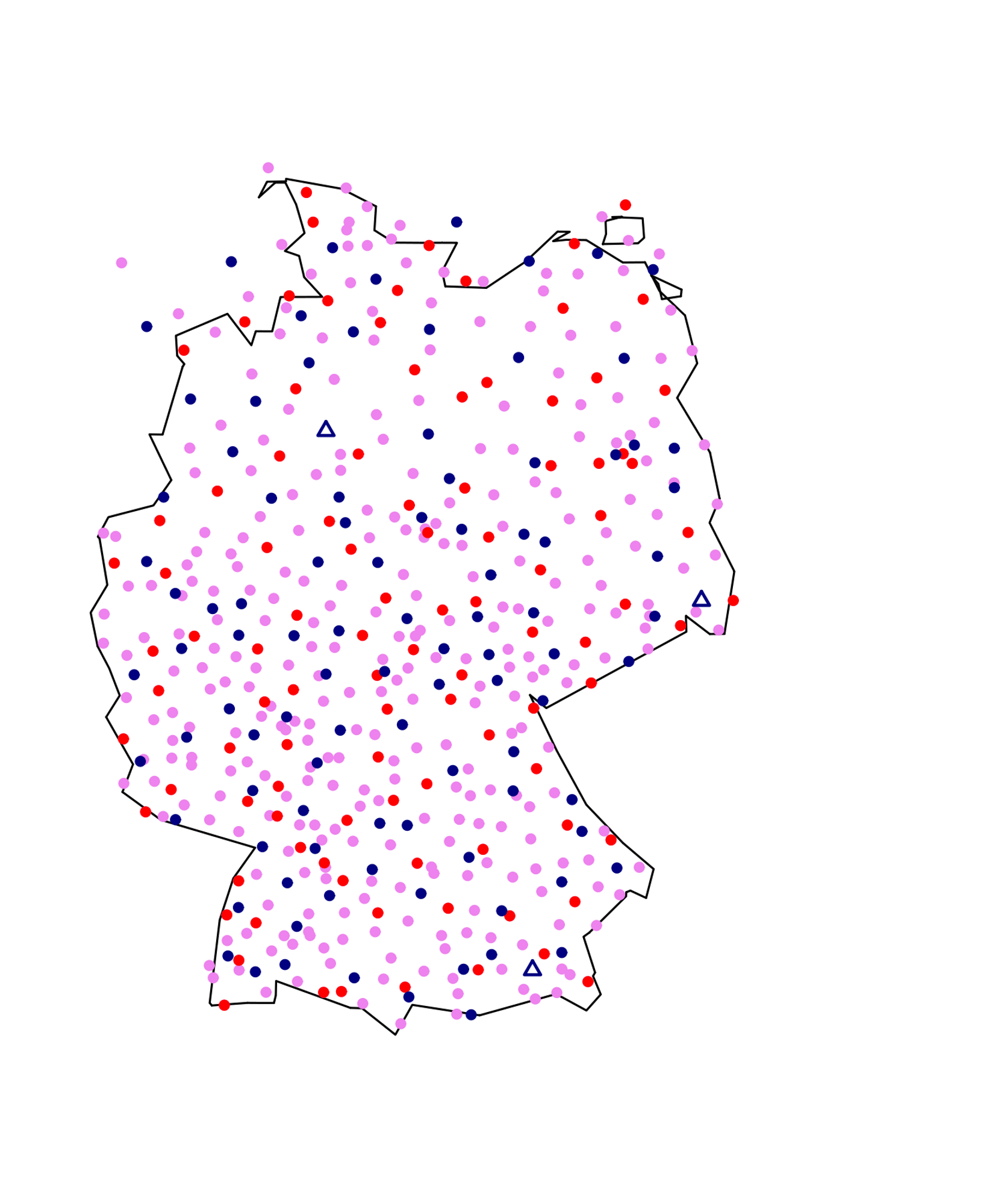}
 \caption{(Smoothed) topography of Germany (left plot) 
 and observation station locations (right plot). Fitting stations are shown by red/violet dots with in-sample validation stations in red. The blue dots are the validations stations that were not used for model fitting, the three blue triangles correspond to the stations at Nienburg, Kubsch\"utz, and Rosenheim which are studied later in detail.}
 \label{Fig:TopoTrainVerif}
\end{figure}

In Section \ref{sec:2} we present and motivate our model and explain how the parameters can be estimated. The optimal interpolation of the temperature averages and the uncertainty predictor variable is then the subject of Section \ref{sec:3}. In Section \ref{sec:4} we use our method and, for comparison, BMA, GMA and EMOS, for post-processing temperature ensemble forecasts from the COSMO-DE-EPS, a newly developed high-resolution EPS for Germany \citep{Gebhardt&2011}. We finish with a discussion and briefly point out further extensions.

\section{Locally adaptive EMOS model}
\label{sec:2}

\begin{figure}
 \centering
 \includegraphics[width=1\textwidth]{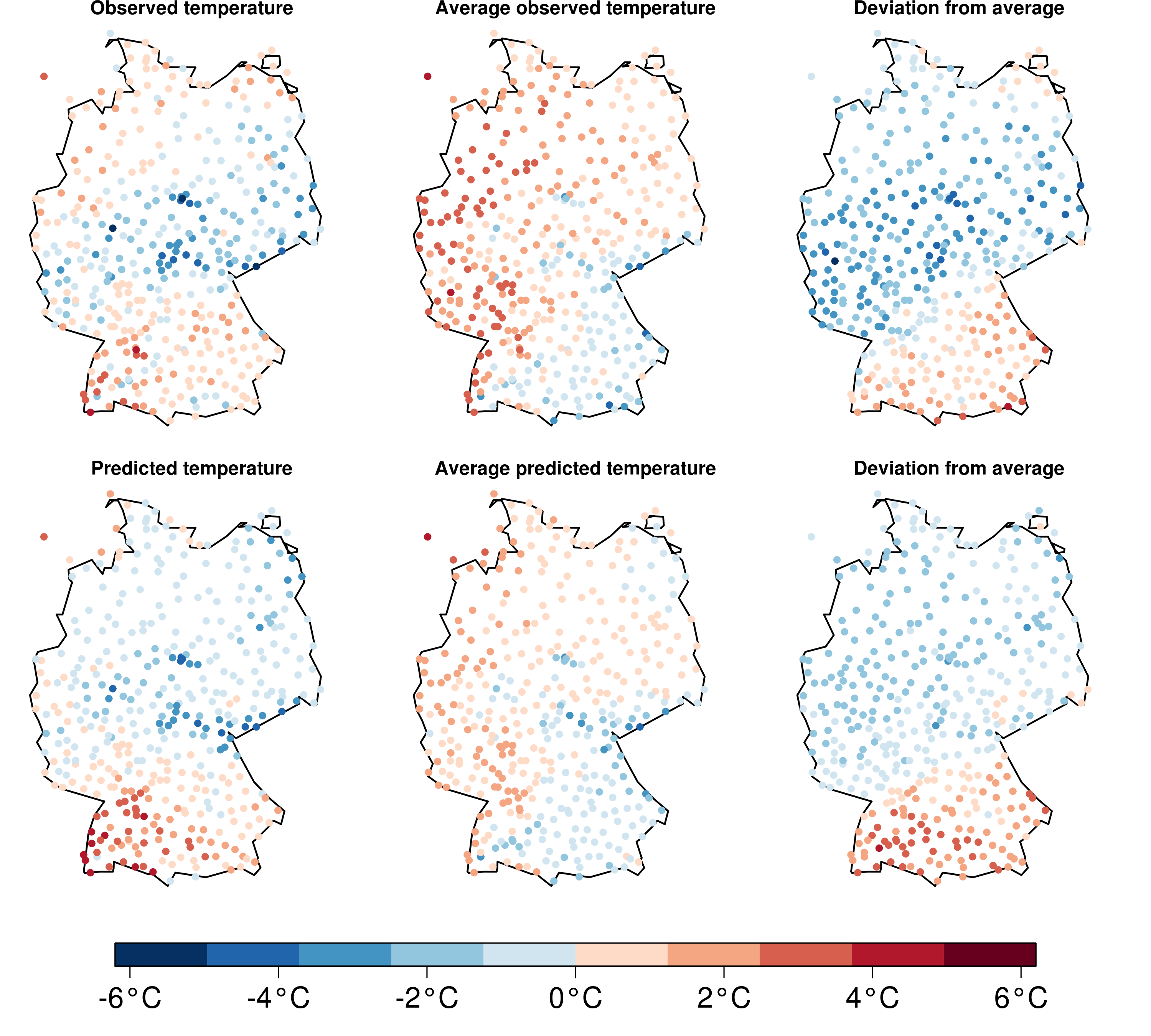}
 \caption{Observed (top row) and predicted (bottom row) temperatures over Germany at 1800 UTC. From left to right the plots show temperatures on January 10, 2011, temperature averages of all days of January 2011, and the difference between the two former.}
 \label{Fig:TempComponents}
\end{figure}

Let $y_{st}$ be the temperature at location $s\in\calS$ and time $t$ and $f_{st1},\ldots,f_{stm}$ the corresponding ensemble member forecasts. Denote by $\calT$ the set of training days, $\#\calT$ its cardinality, and define
\[
 \bar{y}_s := \frac{1}{\#\calT}\sum_{t\in\calT} y_{st}, \quad \bar{f}_{sk} := \frac{1}{\#\calT}\sum_{t\in\calT} f_{stk}, \quad k=1,\ldots,m.
\]
Our basic model is then
\begin{equation}\label{Eq:BasicModel}
 y_{st} = \bar{y}_s + b_1(f_{st1}-\bar{f}_{s1}) + \ldots +  b_m(f_{stm}-\bar{f}_{sm}) + \varepsilon_{st}, \qquad \varepsilon_{st} \sim \calN(0,\sigma_s^2),
\end{equation}
where $b_1,\ldots,b_m$ are regression coefficients and $\sigma_s^2$ is the forecast uncertainty. It differs from the NGR model by \citet{Gneiting&2005} in that it does not include an additive and multiplicative bias correction, but takes the form of the conditional expectation of observations given the forecasts under the assumption of a multivariate normal distribution.
Even though the parameters $b_1,\ldots,b_m$ are not site-specific, our model is locally adaptive as a consequence of the centering of both forecasts and observations. To illustrate this point, we already take a look at the data that will be analyzed in Section \ref{sec:4}. We consider ensemble temperature forecasts over Germany at 1800 UTC by the COMSO-DE-EPS (details are given in Section \ref{sec:4}) initialized at 0000 UTC. The forecasts are calibrated against temperature measurements at 404 SYNOP stations over Germany (red and violet dots in Figure \ref{Fig:TopoTrainVerif}). In Figure \ref{Fig:TempComponents} we depict the variables $y_{st}, \bar{y}_s$ and $y_{st}-\bar{y}_s$, as well as $f_{st1},\bar{f}_{s1}$ and $f_{st1}-\bar{f}_{s1}$ for a specific day/training period. It appears that a good deal of the small scale spatial variability of $y_{st}$ can be attributed to the mean component $\bar{y}_s$, while the deviation $y_{st}-\bar{y}_s$ from this mean is comparatively smooth. The same is true for the 
forecasts, except for the fact that the spatial variability of $\bar{f}_{s1}$ is lower than that of $\bar{y}_s$. This is a consequence of incompletely resolved orography and other sources of variability that cannot be considered by the NWP model. It is therefore mainly the inclusion of $\bar{y}_s$ in \eqref{Eq:BasicModel} that accounts for site-specific temperature anomalies, and resolves variations that cannot be resolved at the model grid scale.

The forecast variance $\sigma_s^2$ is represented as a sum $\sigma_s^2=c_1\xi_s^2+c_2S_s^2$ that includes the ensemble variance $S_s^2$ at location $s$ and a further predictor variable $\xi_s^2$ reflecting the local uncertainty during the training period. This variable is obtained as follows: let $f_{st\star}$ be the mean of the ensemble member forecasts and denote by $\bar{f}_{s\star}$ the corresponding average over $\calT$. Then define $\xi_s^2$ as the mean of the squared residuals at location $s$ of a simplified regression model with the centered ensemble mean as the only predictor:
\[
 \xi_s^2 := \frac{1}{\#\calT} \sum_{t\in\calT} \big( y_{st} - \bar{y}_s - \hat{b}_\star(f_{st\star}-\bar{f}_{s\star}) \big)^2,
 \qquad \hat{b}_\star = \frac{\sum_{t\in\calT}\sum_{\tilde{s}\in\calS_{obs}} (y_{\tilde{s}t}-\bar{y}_{\tilde{s}})(f_{\tilde{s}t\star}-\bar{f}_{\tilde{s}\star})}{\sum_{t\in\calT}\sum_{\tilde{s}\in\calS_{obs}}(f_{\tilde{s}t\star}-\bar{f}_{\tilde{s}\star})^2}
\]
The reason for using a simplified rather than the full regression model is that the $\xi_s^2$ is essentially a preliminary estimate of the local forecast variance. Its overall magnitude is not important at this point, but we want our preliminary variance estimate to be as stable as possible, and this can usually be better achieved with parsimonious parametrizations. Given $\xi_s^2$, we can proceed in the same way as \citet{Gneiting&2005} and choose the model parameters $b_1,\ldots,b_m,c_1,c_2$ as the minimizers of the continuous ranked probability score \citetext{CRPS, \citealp{Hersbach2000}} of the respective predictive distributions on the training data. Such minimum score estimation is quite natural since proper scoring rules honor probabilistic forecasts which attain a good compromise between sharpness and calibration \citep{GneitingRaftery2007}, and it permits constraining the parameters to positive values so that the regression coefficients $b_1,\ldots,b_m$ can be interpreted as weights.
The CRPS for a Gaussian distribution can be expressed in closed form, making minimum CRPS estimation computationally rather efficient unless $m$ is very big. In this case, there is also a certain danger of overfitting, but both issues could be accounted for by identifying subgroups of ensemble members for which it makes sense to assume identical coefficients $b_i$. The same has to be done if certain ensemble members are {\em exchangeable}.


\section{Spatial interpolation of $\mathbf{\bar{y}_s}$ and $\mathbf{\xi_s^2}$}
\label{sec:3}

The procedure described so far can be used to fit an NGR model and obtain predictive distributions at sites $s\in\calS_{obs}$ by plugging the new ensemble forecasts ($\bar{f}_{s1},\ldots,\bar{f}_{sm}$ remain unchanged) and the ensemble variance into \eqref{Eq:BasicModel}. While $b_1,\ldots,b_m,c_1,c_2$ are global parameters, the temperature averages $\bar{y}_s$ and our predictor $\xi_s^2$ for the local uncertainty have to be extrapolated to locations $s\notin\calS_{obs}$ where predictive distributions are desired. Due to their considerable variability, spatial interpolation of these variables is a challenging task, especially for $\bar{y}_s$. On the one hand, there are complicated large scale temperature gradients which are hard to capture by a deterministic trend function. On the other hand, we observe a number of strong local deviations from the large scale trends, which are rather irregularly distributed in space and change over time, although in some regions such irregularities are much more common than 
in others. Our interpolation scheme is based on the following modeling assumptions
\begin{itemize}
 \item we consider $\bar{y}_s, s\in\calS$ as a realization of an intrinsic Gaussian random field $\{\bar{Y}_s:s\in\calS\}$ with generalized covariance function
 \[
 \Cov\big(\bar{Y}_s,\bar{Y}_{\tilde{s}}\big) = \theta_{y,1}\cdot\zeta_{y,s}\cdot\indicator{s=\tilde{s}} - \theta_{y,2}\cdot\|s-\tilde{s}\| =: C_y(s,\tilde{s}),
 \]
 \item we consider $z_s:=2\log(\xi_s), s\in\calS$ as a realization of an intrinsic Gaussian random field $\{Z_s:s\in\calS\}$  with generalized covariance function
 \[
 \Cov\big(Z_s,Z_{\tilde{s}}\big) = \theta_{z,1}\cdot\zeta_{z,s}\cdot\indicator{s=\tilde{s}} - \theta_{z,2}\cdot\|s-\tilde{s}\| =: C_z(s,\tilde{s}),
 \]
\end{itemize}
where $\indicator{s=\tilde{s}}$ denotes the indicator function that is $1$ if $s=\tilde{s}$ and otherwise $0$, and $\|s-\tilde{s}\|$ is the distance between $s$ and $\tilde{s}$. The concept of an {\em intrinsic} random field implies that neither the means nor the variances and covariances of the random variables $Z(s),Z(\tilde{s})$ are defined. Instead, the dependence structure of $Z$ (analogously for $\bar{Y}$) is specified only through linear combinations of the form
\begin{equation}\label{Eq:AllowableLC}
 Z_\lambda:=\sum_{i=1}^n \lambda_iZ(s_i), \quad \mbox{ with } \lambda_1,\ldots,\lambda_n \mbox{ such that } \; \sum_{i=1}^n \lambda_i\,p(s_i)=0 \; \mbox{ for all } p\in\calP,
\end{equation}
where $\calP$ is some linear function space. All of these {\em allowable} linear combinations have mean zero by definition, and
\[
 \quad \E(Z_\lambda Z_\lambda) = \sum_{i=1}^n\sum_{j=1}^n \lambda_i\lambda_j C_z(s_i,s_j)
\]
so that $C_z$ and $C_y$ indeed assume the role of a covariance function but are restricted to allowable linear combinations. They generalize the class of covariance functions because they are only required to be 
{\em conditionally positive definite} with respect to $\calP$. Our specific choice implies that $\bar{Y}$ and $Z$ are Brownian surfaces with a site-specific nugget effect, and $C_z$ and $C_y$ are conditionally positive definite if $\calP$ contains the constant functions. Realizations of Brownian surfaces locally behave like those of an exponential covariance model, but they have no tendency to revert to some mean, and are therefore well suited to model irregular large scale fluctuations \citetext{see \citealp{chiles-delfiner}, Ch.~4 for a very nice motivation and further details on intrinsic random functions}.  

In virtually all instances in the statistical literature where intrinsic random fields are used for spatial modeling, $\calP$ is chosen such as to ensure conditional positive definiteness of the desired generalized covariance function. For the interpolation of $Z$ we do the same and let $\calP$ consist of constant function only. For $\bar{Y}$, however, we let $\calP$ be the span of the three B-spline basis functions of the altitude $a(s)$ at $s$ that correspond to a natural cubic spline with boundary knots at $0$m and $1500$m and an interior knot at $1000$m. When intrinsic kriging \citep{matheron-1973} of $\bar{Y}$ is performed, condition \eqref{Eq:AllowableLC} forces the kriging weights to be consistent with any vertical temperature profile that can be represented by a function from $\calP$. Choosing this space larger than required for technical reasons therefore allows us to account for the decrease of temperature with altitude. Our particular choice is a compromise between flexibility -- even for average 
temperatures we sometimes observe inversions and our model should be able to deal with them -- and availability of a sufficient amount of data at high altitudes. The corresponding Kriging system is solvable if the three B-spline basis functions, considered as functions in $s$, are linearly independent on the set $\calS_{obs}$ \citetext{for mathematical details see also \citealp{scheuerer-et-al-2012-EJAM}}.

The preceding explanations mostly refer to the components of $C_y$ and $C_z$ which model the large scale fluctuations. Small scale fluctuations of $\bar{Y}$ and $Z$ are accounted for by a site-specific nugget effect, and we still need to explain how we define the variables $\zeta_{y,s}$ and $\zeta_{w,s}$. Our reason for not using a uniform nugget effect is that even after considering altitude-related effects, some individual stations strongly differ from their neighborhood, while on the other hand large regions (e.g.\ the North German Plain) appear rather homogeneous with no apparent small scale variability. As an indicator for some station's dissimilarity with its neighborhood we use the squared, standardized leave-one-out cross validation (LOOCV) errors with respect to a Brownian surface {\em without} nugget effect. Let $p_1,\ldots,p_k$ be a basis of $\calP$, $\calS_{obs}=\{s_1,\ldots,s_n\}$, $\bar{\mathbf{y}}=(\bar{y}_{s_1},\ldots,\bar{y}_{s_1})^\prime$ and define
\[
 A := - \left( \begin{array}{cccc}
       0 & \|s_1-s_2\| & \ldots & \|s_1-s_n\| \\
      \vdots & & & \vdots \\
       \|s_n-s_1\| & \ldots & \|s_n-s_{n-1}\| & 0 \\
      \end{array} \right), \;
 P := \left( \begin{array}{ccc}
       p_1(s_1) & \ldots & p_k(s_1) \\
      \vdots & & \vdots \\
       p_1(s_n) & \ldots & p_k(s_n) \\
      \end{array} \right)
\]
Denote by $e_i$ the $i^{th}$ LOOCV error, i.e.\ the difference between the true and the interpolated value at $s_i$ based on the values at all remaining observational sites. A straightforward generalization of the arguments in \citet{rippa-1999} shows that
\[
 e_i = \frac{\alpha_i}{\Psi_{ii}} , \; i=1,\ldots,n, \quad \mbox{where} \quad
    \left( \begin{array}{cc}
       \Psi & \ast \\
       \ast & \ast \\
      \end{array} \right)
  := \left( \begin{array}{cc}
       A & P \\
       P^\prime & 0 \\
      \end{array} \right)^{-1}
    \quad \mbox{and} \quad
    \alpha = \Psi\bar{\mathbf{y}}.
\]
Moreover, the kriging variance (the expected squared interpolation error) of the leave-one-out interpolation at $s_i$ is equal to $1/\Psi_{ii}$, which leads us to define $\tilde{\zeta}_{y,i}=\alpha_i^2/\Psi_{ii}, \: i=1,\ldots,n$. A large value of $\tilde{\zeta}_{y,i}$ means that $\bar{y}_{s_i}$ deviates from its interpolant much stronger than one would expect, and we therefore assume a bigger nugget effect at $s_i$ than at locations where the squared, standardized LOOCV error is small. Since our definition of $A$ does not involve any variance parameter, the absolute values of $\tilde{\zeta}_{y,1},\ldots,\tilde{\zeta}_{y,n}$, are meaningless, but their relative differences help distinguish atypical stations or regions from those that seem to agree well among each other. We proceed in exactly the same way to define $\tilde{\zeta}_{z,1},\ldots,\tilde{\zeta}_{z,n}$ with the only difference being that $P$ is now simply a vector of length $n$ with all elements equal to $1$. The squared, standardized LOOCV errors 
are a simple but useful and easy-to-calculate indicator for local anomalies, but they are subject to sampling variability and should therefore be smoothed, even if this implies again a loss in spatial resolution. We define the final predictors
\[
 \zeta_{y,s} := \sum_{i=1}^n w_i(s) \: \tilde{\zeta}_{y,i}, \quad
 \zeta_{z,s} := \sum_{i=1}^n w_i(s) \: \tilde{\zeta}_{z,i},
 \qquad w_i(s) = \frac{K(\|s-s_i\|/\lambda_s)}{\sum_{j=1}^n K(\|s-s_j\|/\lambda_s)},
\]
where $K$ is a smoothing kernel and $\lambda_s$ is a location-dependent bandwidth parameter. We choose $K(h)=\indicator{h<1}(1-h^2)^3$ (``triweight kernel'') which yields smooth spatial transitions while the compact support speeds up the computation time for the above sums. The bandwidth parameter $\lambda_s$ is chosen as the distance of $s$ to the $25$th nearest station.
For a complete specification of $C_y$ and $C_z$ we still need to estimate the parameters $\theta_{y,1},\theta_{y,2}$ and $\theta_{z,1},\theta_{z,2}$, respectively. This can be done via {\em restricted maximum likelihood estimation}, an extension of the standard maximum likelihood approach that also works for intrinsic random fields \citetext{see e.g.~\citealp{stein}, Sec.~6.4}.

\begin{figure}
 \centering
 \includegraphics[width=1\textwidth]{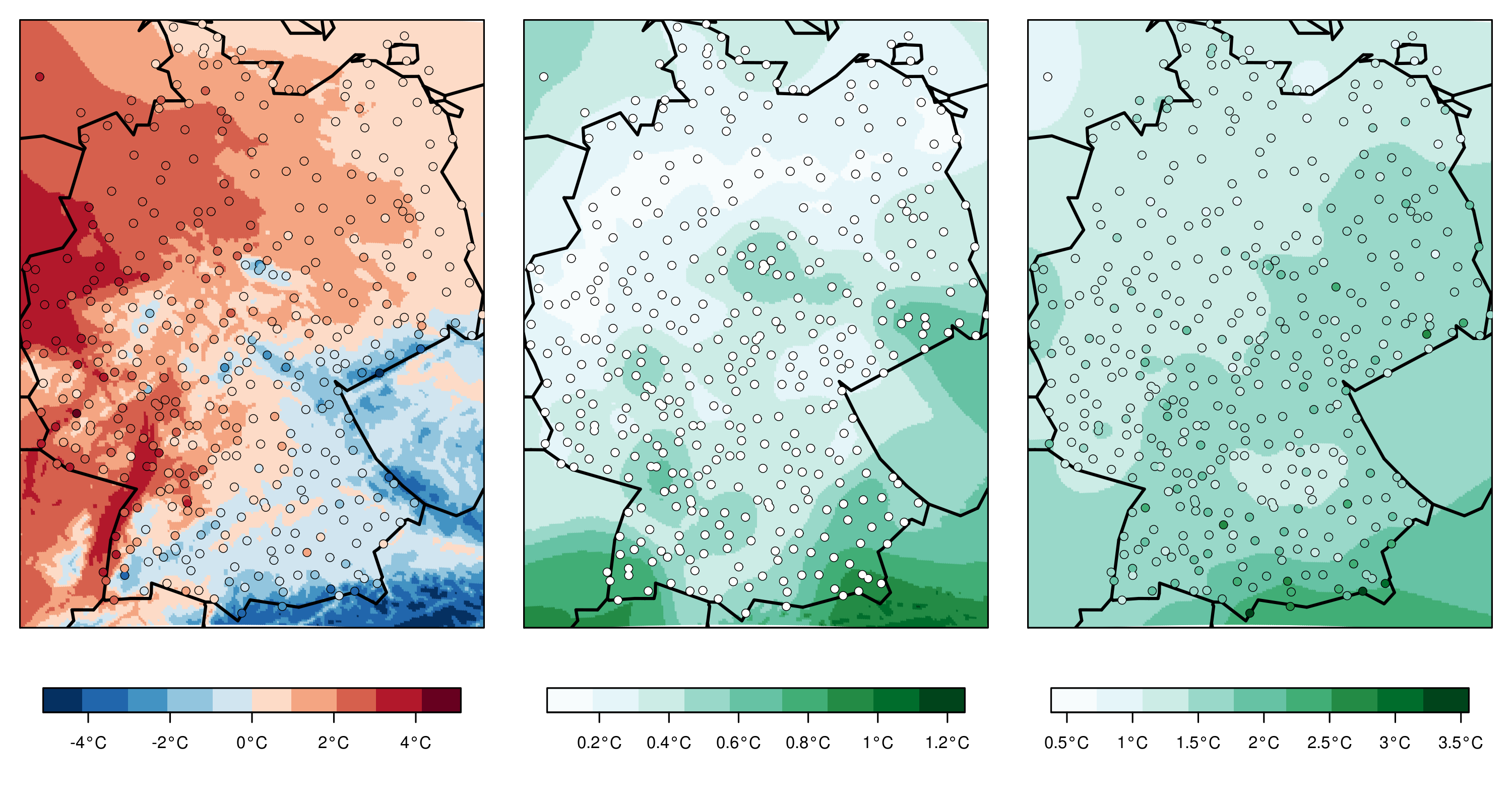}
 \caption{Interpolated average temperatures at 1800 UTC in January 2011 (left), the corresponding kriging standard deviation (middle) and interpolated predictor variable $\xi_s$ for the forecast uncertainty (right).}
 \label{Fig:InterpolationExample}
\end{figure}

With all predictors and covariance parameters been defined, we can set up and solve the intrinsic kriging systems and compute the interpolants of $\bar{y}_s, s\in\calS_{obs}$ and $z_s,s\in\calS_{obs}$. The interpolation uncertainty in both cases can be quite large, and at least in the case of the average temperatures it can be accounted for in a straightforward way by calculating the kriging variance $\sigma_{\bar{y},s}^2$, and adding it to the forecast uncertainty $\sigma_s^2$. To illustrate how the interpolation procedures discussed in this section work with real data, we depict examples of interpolated average temperatures $\bar{y}_s, s\in\calS_{grid}$, the corresponding kriging standard deviations $\sigma_{\bar{y},s}, s\in\calS_{grid}$ and the interpolated predictor variable $\xi_s, s\in\calS_{grid}$ for the forecast uncertainty in Figure \ref{Fig:InterpolationExample}. Although the extrapolation of $\bar{y}_s$ beyond the German borders should not be considered very reliable, one can clearly see how the 
inclusion of the B-spline basis functions in $\calP$ permit a certain extrapolation even to those altitudes where no observations are available (see also the topographical map in Figure \ref{Fig:TopoTrainVerif}). The kriging standard deviations are dominated by the location dependent nugget effects, which lead to very strong smoothing and high uncertainty in regions with strong differences between neighboring stations that can not (or not entirely) be explained by their altitude differences. The predictor variable $\xi_s$ is smoothed quite strongly, and one cannot hope to accommodate the small scale variations of forecast uncertainty. Yet, it captures some large scale spatial variations and can be expected to improve on the forecast uncertainty model from the standard EMOS approach which fully relies on the ensemble variance as a predictor for spatial differences.

\section{Data example}
\label{sec:4}

We use the postprocessing method described above to generate predictive distributions for temperature over Germany at 1800 UTC based on ensemble forecasts by the COMSO-DE-EPS. This is a multi-analysis and multi-physics ensemble prediction system based on the high-resolution numerical weather prediction model COSMO-DE \citep{Baldauf&2011}, a configuration of the COSMO model with a horizontal grid size of 2.8 km operated by the German Meteorological Service (DWD). It is pre-operational since 9 December 2010, covers the area of Germany and produces forecasts with lead times up to 21 hours. The current setup of the lateral boundary conditions uses forecasts of different global models, while different configurations of the COSMO-DE model are used for the variation of model physics \citetext{for further details see \citealp{Gebhardt&2011}}. A new model run is started every three hours, we use the one initialized at 0000 UTC which corresponds to a lead time of 18 hours. The COSMO model uses a rotated spherical 
coordinate system in order to project the geographical coordinates to the plane with distortions as small as possible \citep[Sec.~3.3]{DomsSchaettler2002}, and we adopt this coordinate system to calculate horizontal distances.\\
Probabilistic forecasts are produced and evaluated during the period from 1 February 2011 to 31 January 2012. Training and verification is performed with temperature observations from 504 SYNOP stations in Germany. Stations with more than 100 missing values during the verification period have been omitted. The station at the summit of ``Zugspitze'', the highest mountain in Germany (2960m) was omitted as well; given that overall only 6 stations are above 1000m with the second highest at 1832m, we found that ``Zugspitze'' has an undesirably high leverage on the adjustment of the vertical temperature profile. From the 504 stations, 100 are left out and used for out-of-sample verification. The remaining 404 stations are used for training, 100 of them are also used for in-sample verification. Figure \ref{Fig:TopoTrainVerif} gives an overview over all stations used for training and verification.

As a benchmark for our postprocessing methods, we use the standard EMOS method by \citet{Gneiting&2005}, the BMA method \citep{Raftery&2005} implemented in the 'ensembleBMA' package in R \citep{Fraley&2011}, and its locally adaptive extension GMA proposed by \citet{Kleiber&a2011}. Unlike BMA, the GMA method only performs an additive bias correction but estimates bias and variance parameters separately for each $s\in\calS_{obs}$ and interpolates them to locations $s\notin\calS_{obs}$ by geostatistical methods similar to the ones we use in our approach. We also experimented with a variant of GMA that performs both additive and multiplicative (site-specific) bias correction. For a training period of $50$ days this variant yielded better results than the original method by \citet{Kleiber&a2011}, but in this case the results of both GMA variants were inferior to those presented below, obtained with a rolling 30 day training period. This choice is similar to the 25 days used by \citet{Raftery&2005} and \citet{
Kleiber&a2011} who found this to be the smallest number of training days for which a good compromise between adaptivity to seasonal changes and precision of parameter estimates is obtained.\\
The variation of model physics in the current implementation of the COSMO-DE-EPS is tailored to represent uncertainty in the precipitation-generating processes. The temperature forecasts of the respective ensemble members are almost identical, and we therefore take their means and work with the four member ensemble that corresponds to the four different boundary conditions. To evaluate the raw ensemble's predictive performance we use of course all 20 members.

\begin{table}
\centering
 \caption{In-sample results for the raw ensemble and the different postprocessing methods: MAE, CRPS and width and empirical coverage of prediction intervals with a nominal coverage of $81\%$ and $90.5\%$, respectively.}
 \label{Tab:InSample}
 \smallskip
 \begin{tabular}{lcccccc}
  \toprule 
  & & & \multicolumn{2}{c}{81\% pred.~int.} & \multicolumn{2}{c}{90.5\% pred.~int.} \\
  & MAE (\textcelsius) & CRPS (\textcelsius) & width (\textcelsius) & coverage (\%) &  width (\textcelsius) & coverage (\%)\\
  \midrule 
 Ensemble    & 1.433 & 1.244 & 1.18 & 25.6 & 1.45 & 31.2 \\
 EMOS        & 1.328 & 0.950 & 4.20 & 79.6 & 5.35 & 88.8 \\
 BMA         & 1.330 & 0.950 & 4.38 & 81.5 & 5.58 & 90.4 \\
 adpt.~EMOS & 1.265 & 0.903 & 3.83 & 77.2 & 4.88 & 87.0 \\
 GMA         & 1.271 & 0.918 & 4.30 & 79.8 & 5.46 & 88.4 \\
  \bottomrule  
 \end{tabular}
\end{table}

\section*{In-sample performance}

To begin with, we assess the in-sample predictive performance, i.e.\ the performance at a subset of the training stations. In these cases, $\bar{y}_s$ is known, $\sigma_s^2$ and the other postprocessing parameters can all be calculated as described in Section \ref{sec:2}, and no interpolation is necessary. The same is true for the local parameters in the GMA approach. Table \ref{Tab:InSample} shows the mean absolute error (MAE) the continuous ranked probability score (CRPS) as well as the width and empirical coverage of prediction intervals with a nominal coverage of $81\%$ and $90.5\%$ for the probabilistic forecasts obtained from the raw ensemble and the different postprocessing methods. $90.5\%$ and $81\%$ are the probabilities that the observation lies within the ensemble range, and the range of all but the lowest and the highest ensemble member forecasts, respectively, assuming that these are independent draws of the same distribution as the one of the observation.
It is apparent from its narrow prediction intervals and the poor coverage that the raw ensemble is strongly underdispersive. All postprocessing methods largely correct this and attain much better scores. Both GMA and our adaptive EMOS method yield a further improvement over their non-adaptive counterparts. Our approach is a bit underdispersive, but still somewhat ahead of GMA with respect to both MAE and CRPS, which suggests that it is more successful in removing local biases. Despite its simplicity, representation \eqref{Eq:BasicModel} of temperature as a sum of a site-specific short-term average and NWP forecast-driven terms seems to constitute an appropriate description of real temperatures. It is not clear, however, if the advantages of our model carry over to the situation where $\bar{y}_s$ is unknown, and has to be found by interpolation. This will be assessed in the following.

\section*{Out-of-sample performance}

Table \ref{Tab:OutOfSample} is the counterpart of Table \ref{Tab:InSample} in the situation where we predict the temperature at locations where we have not used the observations for training. For the non-adaptive methods this makes only little difference since they do not rely on site-specific information anyway, while the adaptive ones have to rely on interpolation to obtain such information. The scores in 
Table \ref{Tab:OutOfSample} show that both GMA and our adaptive EMOS approach loose a bit of their advantage over BMA and standard EMOS, but still present a significant improvement with substantially narrower prediction intervals but just slightly inferior coverage. Note that the width of the adaptive EMOS prediction intervals increases compared to the in-sample case as a result of the additional uncertainty due to the interpolation of $\bar{y}_s$, which we explicitly consider in our model. Overall, the interpolation scheme suggested in Section \ref{sec:3} seems capable to carry over the advantages of our model noted above to the out-of-sample situation which one faces in practice.

\begin{table}
\centering
 \caption{Out-of-sample results for the raw ensemble and the different postprocessing methods: MAE, CRPS and width and empirical coverage of prediction intervals with a nominal coverage of $81\%$ and $90.5\%$, respectively.}
 \label{Tab:OutOfSample}
 \smallskip
 \begin{tabular}{lcccccc}
 \toprule 
 & & & \multicolumn{2}{c}{81\% pred.~int.} & \multicolumn{2}{c}{90.5\% pred.~int.} \\
  & MAE (\textcelsius) & CRPS (\textcelsius) & width (\textcelsius) & coverage (\%) &  width (\textcelsius) & coverage (\%)\\
 \midrule 
 Ensemble    & 1.436 & 1.250 & 1.16 & 25.0 & 1.43 & 30.4 \\
 EMOS        & 1.333 & 0.951 & 4.19 & 79.4 & 5.34 & 88.7 \\
 BMA         & 1.337 & 0.951 & 4.38 & 81.4 & 5.58 & 90.2 \\
 adpt.~EMOS & 1.313 & 0.937 & 4.03 & 77.8 & 5.14 & 87.7 \\
 GMA         & 1.317 & 0.943 & 4.17 & 78.5 & 5.30 & 87.7 \\
 \bottomrule  
 \end{tabular}
\end{table}

\begin{figure}
 \centering
 \includegraphics[width=1\textwidth]{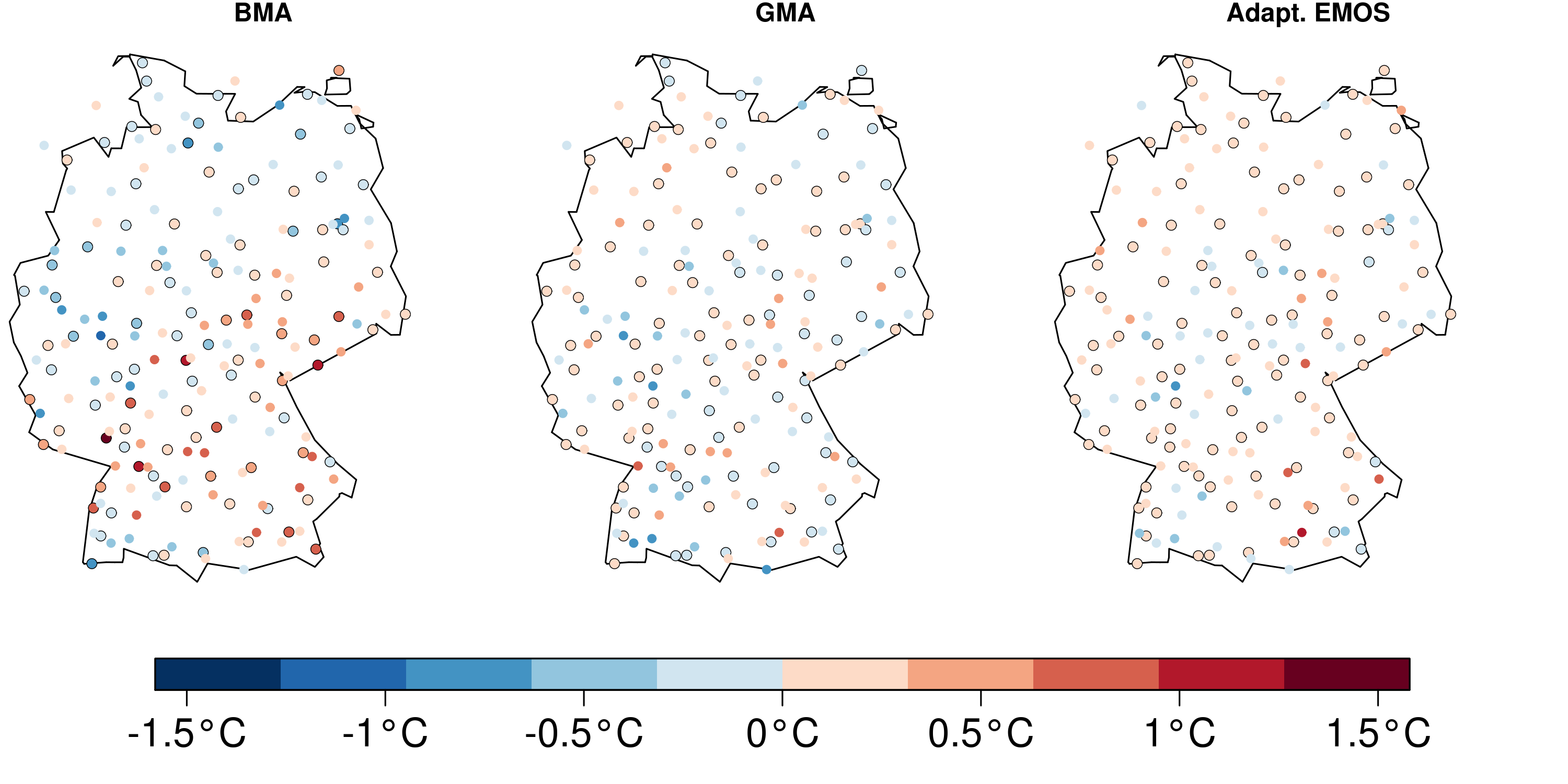}
 \caption{Annual means of the forecast errors of BMA (left), GMA (middle) and our adaptive EMOS method (right). The black circles mark the 100 in-sample validation stations, the remaining 100 points are the out-of-sample validation stations.}
 \label{Fig:ForecastBiases}
\end{figure}

To illustrate the additional challenges of out-of-sample prediction and see where the adaptive methods could be further improved, we depict the mean of the local forecast errors of BMA, GMA, and adaptive EMOS over the whole verification period in Figure \ref{Fig:ForecastBiases}. The local biases revealed this way are rather strong and frequent for BMA, which does not account for local particularities. Both GMA and our proposed EMOS method reduce these local biases dramatically: there are virtually no biases left at the in-sample stations and biases are strongly reduced at most of the out-of-sample stations, although there are still a few stations left were biases persist. These exceptional cases are not always the same for GMA and our method, which shows that the ways of striving for adaptivity are really qualitatively different. With our method, the strongest under-prediction of temperature occurs at station L841 at Frankfurt/Main, while the strongest over-prediction occurs at station P830 at Oberhaching 
which is located some 15 km south of Munich city. A possible explanation for this might be warmer city climate of Frankfurt and Munich: that of Frankfurt is not anticipated by the surrounding stations while that of Munich is wrongly extrapolated to its surrounding. This suggests that appropriate incorporation of land use information in our interpolation scheme might help explain at least part of the small scale variability. It may even make the predictor variables $\zeta_{y,s}$ and $\zeta_{z,s}$ redundant, and explain local anomalies rather than just mitigating their effects on interpolation.

We finally take a closer look at a few specific stations to illustrate why not only forecast biases, but also forecast uncertainties should be modeled in a locally adaptive way. Table \ref{Tab:CaseStudies} gives the average width and empirical coverage of prediction intervals with a nominal coverage of $81\%$ for the stations at Nienburg/Weser (North German Plain), Kubsch\"utz (Lusatian Highlands) and Rosenheim (Alpine foothills). The average widths of the non-adaptive methods differ just slightly among the different stations. Their empirical coverages, on the contrary, exceed the nominal coverage at Nienburg but are too low at Rosenheim, suggesting that temperature prediction is more difficult in mountainous terrain. Both adaptive methods accommodate this fact and issue wider prediction intervals at Kubsch\"utz and Rosenheim on the one hand, and narrower prediction intervals at Nienburg on the other hand, which results in empirical coverages that are generally much closer to the nominal coverage. The 
adaptive EMOS method presented in this paper is overall slightly underdispersive but mostly achieves a locally adequate representation of forecast uncertainty.
 
\begin{table}
\centering
 \caption{Average width and empirical coverage of prediction intervals (nominal coverage $81\%$) for temperature at three selected stations.}
 \label{Tab:CaseStudies}
 \smallskip
 \begin{tabular}{lcccccc}
 \toprule 
 & \multicolumn{2}{c}{Nienburg/Weser} & \multicolumn{2}{c}{Kubsch\"utz} & \multicolumn{2}{c}{Rosenheim} \\
 & width (\textcelsius) & coverage (\%) & width (\textcelsius) & coverage (\%) &  width (\textcelsius) & coverage (\%) \\
 \midrule 
 Ensemble    & 1.20 & 31.0 & 1.02 & 19.2 & 1.27 & 21.9 \\
 EMOS        & 4.23 & 83.8 & 4.16 & 77.6 & 4.18 & 73.7 \\
 BMA         & 4.39 & 85.1 & 4.36 & 81.3 & 4.37 & 75.2 \\
 adapt.~EMOS & 3.73 & 80.0 & 4.25 & 80.0 & 4.80 & 79.2 \\
 GMA         & 3.65 & 79.1 & 4.43 & 81.8 & 4.59 & 75.8 \\
 \bottomrule  
 \end{tabular}
\end{table}

\section{Discussion}
\label{sec:5}

The adaptive EMOS method presented in this article generalizes the approach by \citet{Gneiting&2005} and generates locally calibrated probabilistic temperature forecasts based on the output of an EPS and observational data that is used for calibration. It represents the mean of the predictive distributions as a sum of short-term averages of local temperatures and EPS-driven terms, which implies a simple and intuitive predictive distribution model with relatively few parameters. The good in-sample results (see Section \ref{sec:4}) suggest that this model is adequate for describing the true nature of temperature forecast uncertainty. In order to generate probabilistic forecast at non-observational locations we have proposed a geostatistical interpolation method based on an intrinsic Gaussian random field model and a simple but effective technique to account for different magnitudes of small scale variability in different regions. This interpolation scheme also permits a realistic assessment of the additional 
uncertainty that originates from the fact that the true average local temperature is unknown. It led to locally calibrated forecasts at most of the out-of-sample validation stations of our data example, and our method overall compared favorably with the GMA approach by \citet{Kleiber&a2011}. A problem that is still apparent with our method is that certain small scale temperature fluctuations, e.g.\ near big cities, cannot be anticipated by our interpolation scheme, resulting in a couple of out-of-sample locations where biases persist. We expect, however, that the conceptual simplicity and good interpretability of our model will permit effective usage of e.g.\ land use information which might help to overcome or at least further reduce remaining biases. This might even make the predictor variables $\zeta_{y,s}$ and $\zeta_{y,s}$ for the degree of small scale variability redundant and resolve the sources of such small scale variability explicitly.\\
An issue that also calls for geostatistical methods but has not been touched in this article is the modeling of spatial correlations between different locations. This is important when the interest is in forecasting spatially aggregated quantities like the minimum or maximum temperature in some region. Approaches along the lines of \citet{Gel&2004} and \citet{Berrocal&2007} could be used to extend our model to a multivariate distribution model which can be used to simulate calibrated and spatially consistent temperature fields from which all composite quantities can be derived.\\
Temperature is, at least from a statistical point of view, one of the easier weather variables because it can be adequately described by a Gaussian distribution. This is particularly helpful in a spatial context where Gaussian random field models are by far the most tractable ones. For non-Gaussian weather variables such as precipitation or wind speeds, additional challenges with spatial modeling and spatial interpolation arise. The work by \citet{Kleiber&b2011} exemplifies for the BMA approach for precipitation how such challenges can be met, and we plan to make similar efforts for the EMOS methods for wind \citep{ThorarinsdottirGneiting2010} and precipitation \citep{scheuerer-2012-QJRMS}, which are conceptually simpler and computationally more efficient than the BMA approaches, and seem to work very well for post-processing ensemble forecasts by the German COSMO-DE-EPS.

\section*{Acknowledgement}
The authors are grateful to Christoph Gebhard, Tilmann Gneiting, and Vanessa Stauch for hints and helpful discussions.\\
They thank Sabrina Bentzien and all members of the COSMO-DE-EPS team of Deutscher Wetterdienst (DWD) for their support with the acquisition of the ensemble forecast data.\\
This work is funded by Deutscher Wetterdienst in Offenbach, Germany, in the framework of the extramural research program.

\bibliographystyle{plainnat}
\bibliography{../MSbib}

\end{document}